\documentclass[pss]{wiley2sp} 
\usepackage{amsmath}

\begin{document}

\title{Phase separation at the magnetic-superconducting
transition in La$_{0.7}$Y$_{0.3}$FeAsO$_{1-x}$F$_{x}$}

\titlerunning{Magnetism and superconductivity in
La$_{0.7}$Y$_{0.3}$FeAsO$_{1-x}$F$_{x}$}

\author{%
  Giacomo Prando\textsuperscript{\Ast,\textsf{\bfseries 1}},
  Samuele Sanna\textsuperscript{\textsf{\bfseries 2}},
  Gianrico Lamura\textsuperscript{\textsf{\bfseries 3}},
  Toni Shiroka\textsuperscript{\textsf{\bfseries 4}},
  Matteo Tropeano\textsuperscript{\textsf{\bfseries 3,5}},
  Andrea Palenzona\textsuperscript{\textsf{\bfseries 3}},
  Hans-Joachim Grafe\textsuperscript{\textsf{\bfseries 1}},
  Bernd B\"uchner\textsuperscript{\textsf{\bfseries 1,6}},
  Pietro Carretta\textsuperscript{\textsf{\bfseries 2}},
  Roberto De Renzi\textsuperscript{\textsf{\bfseries 7}}%
}

\authorrunning{G. Prando et al.}

\mail{e-mail
  \textsf{g.prando@ifw-dresden.de}, Phone:
  +49-351-4659-668, Fax: +49-351-4659-540%
}

\institute{%
  \textsuperscript{1}\,Leibniz-Institut
  f\"ur Festk\"orper- und Werkstoffforschung (IFW)
  Dresden, D-01171 Dresden, Germany\\
  \textsuperscript{2}\,Dipartimento di Fisica
  and Unit\`a CNISM di Pavia, Universit\`a di Pavia, I-27100
  Pavia, Italy\\
  \textsuperscript{3}\,CNR-SPIN and Universit\`a di Genova,
  I-16146 Genova, Italy\\
  \textsuperscript{4}\,Laboratorium f\"ur Festk\"orperphysik,
  ETH-H\"onggerberg, CH-8093 Z\"urich, Switzerland\\
  \textsuperscript{5}\,Columbus Superconductors S. p. A.,
  I-16133 Genova, Italy\\
  \textsuperscript{6}\,Institut f\"ur Festk\"orperphysik,
  Technische Universit\"at Dresden, D-01062 Dresden, Germany\\
  \textsuperscript{7}\,Dipartimento di Fisica
  and Unit\`a CNISM di Parma, Universit\`a di Parma, I-43124
  Parma, Italy%
}



\abstract{%
%
%
%
\abstcol{%
  In this paper we report a detailed $\mu^{+}$SR and {}$^{19}$F-NMR
  study of the La$_{0.7}$Y$_{0.3}$FeAsO$_{1-x}$F$_{x}$ class of
  materials. Here, the diamagnetic La$_{1-y}$Y$_{y}$ substitution
  increases chemical pressure and, accordingly, sizeably enhances
  the optimal superconducting transition temperature. We investigate
  the magnetic-superconducting phase transition by keeping
  the Y content constant ($y = 0.3$) and by varying the F content
  in the range $0.025 \leq x \leq 0.15$.}%
  {Our results show how magnetism and superconductivity coexist for
  $x = 0.065$. Such coexistence is due to segregation
  of the two phases in macroscopic regions, resembling what was
  observed in LaFeAsO$_{1-x}$F$_{x}$ materials under applied
  hydrostatic pressure. This scenario is qualitatively different
  from the nanoscopic coexistence of the two order parameters
  observed when La is fully substituted by magnetic rare-earth ions
  like Sm or Ce.}%
}

%
%

\maketitle   

\section{Introduction}
The emergence of high-temperature superconductivity nearby the
disruption of a long-range magnetic state is a common feature of
both the cuprates and iron-pnictide compounds, possibly hinting at a
common magnetic mechanism for the Cooper pairing. In many of the
latter materials, the magnetic and superconducting states are known
to coexist nanoscopically \cite{San10,San11,Wie11,Shi11,She12} while
in others they are reported to be macroscopically segregated
\cite{Jul09,Tex12}. In the REFeAsO$_{1-x}$F$_{x}$ family with RE =
La no coexistence at all has been found up to now \cite{Lue09}, even
if a macroscopic segregation was observed upon the application of
hydrostatic pressure \cite{Kha11}.

In order to better understand this phenomenology in the case of
REFeAsO$_{1-x}$F$_{x}$ (RE1111) with RE = La and to eventually
compare the effect of chemical and hydrostatic pressures in such
materials, we have studied the class of compounds
La$_{0.7}$Y$_{0.3}$FeAsO$_{1-x}$F$_{x}$ at different charge doping
levels $x$. The investigated samples were loose powders prepared as
reported in previous works \cite{Tro09,Mar09}. Here, as an effect of
the internal pressure triggered by the different ionic radii of La
and Y, the diamagnetic La$_{1-y}$Y$_{y}$ substitution enhances the
superconducting transition temperature $T_{\textrm{c}}$ from $26$ K
($y = 0$) to $32$ K ($y = 0.3$) at optimal F doping, namely $x =
0.15$ \cite{Tro09}. The magnetic-superconducting transition was then
investigated by tuning the F content in the range $0.025 \leq x \leq
0.15$ and by keeping the yttrium content fixed to $y = 0.3$.

\section{Spin-density wave phase}

Zero-magnetic field (ZF) $\mu^{+}$SR measurements were performed in
order to follow the $x$-dependence of the magnetic volume fraction
$V_{\textrm{m}}(T)$ of the spin-density wave (SDW) phase.
Experiments were carried out at the GPS facility of the S$\mu$S muon
source at the Paul Scherrer Institut (PSI). For all the investigated
temperature values $\left(T\right)$ the general expression
\begin{eqnarray}\label{EqGeneralFittingZF}
    A_{T}(t) & = & A_{0} \left[1 - V_{\textrm{m}}(T)\right]
    e^{-\frac{\sigma^{2} t^{2}}{2}} + {}\\
    & + & A_{0} \left[a^{\perp}(T) F(t) D^{\perp}(t)
    + a^{\parallel}(T) D^{\parallel}(t)\right]\nonumber
\end{eqnarray}
fits the muon-spin depolarization function as a function of time
$\left(t\right)$. In the paramagnetic limit, $V_{\textrm{m}}(T) =
0$, no static field of electronic origin contributes to the
depolarization and only the weak contribution from the nuclear
magnetic moments leads to a slow gaussian depolarization with
characteristic rate $\sigma$. Below the magnetic-order transition
temperature $T_{\textrm{N}}$, the superscript $\perp$ ($\parallel$)
refers to the fraction of muons experiencing a local static magnetic
field in a perpendicular (parallel) direction with respect to the
initial muon spin polarization. Accordingly, the amplitudes
$a^{\perp,\parallel}$ must satisfy the requirement
$\left[a^{\perp}(T) + a^{\parallel}(T)\right] = V_{m}(T)$. A
coherent precession of the implanted muons around the local magnetic
field $B_{\mu}$ can then be discerned in the $a^{\perp}$ amplitudes
and described by the oscillating function $F(t)$ damped by the
function $D^{\perp}(t) = \exp\left(-\lambda^{\perp}t\right)$.

\begin{figure}[t!]
\vspace{6.2cm} \includegraphics{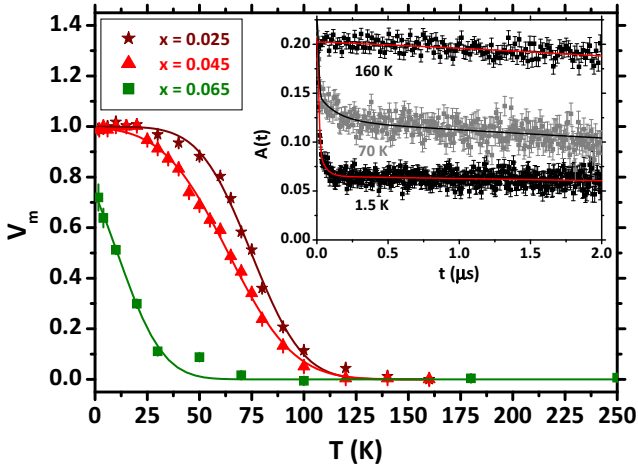} \caption{\label{MagnVolMuSR} Main panel:
magnetic volume fraction $V_{\textrm{m}}(T)$ for $0.025 \leq x \leq
0.065$ after fitting experimental data by means of Eq.
\ref{EqGeneralFittingZF}. Continuous lines are best fits to data
according to erf-like functions. Inset: raw data for $x = 0.025$ at
different $T$ values. Continuous lines are best fits to data
according to Eq. \ref{EqGeneralFittingZF}.}
\end{figure}

If the distribution of $B_{\mu}$ values is too broad, the fast
dephasing of the muon-spins' leads to an overdamping of the signal
preventing one from observing any precession. This is the case for
the investigated samples (see the inset of Fig. \ref{MagnVolMuSR},
relative to the case of $x = 0.025$). It should be remarked that a
small transverse component $a_{\textrm{s}}^{\perp}$ was needed in
addition to the main one, $a_{\textrm{f}}^{\perp}$. The two
subscripts refer to the different extents of slow
($\lambda_{\textrm{s}}^{\perp} \sim 10 \; \mu$s$^{-1}$) and fast
($\lambda_{\textrm{f}}^{\perp} \sim 60 \; \mu$s$^{-1}$) transversal
damping. These two components are likely to be associated with the
two muon sites observed in pure LaFeAsO \cite{DeR12}.

The comparison of $V_{\textrm{m}}(T)$ for $0.025 \leq x \leq 0.065$
is displayed in Fig. \ref{MagnVolMuSR}. It is clear that the effect
of O$^{2-}$/F$^{-}$ substitution is a gradual suppression of the
magnetic critical temperature $T_{\textrm{N}}$, defined as the
temperature where $V_{\textrm{m}}(T) = 0.5$. This strongly resembles
previous reports for other RE1111 materials \cite{Shi11,Lue09}.
Samples are fully magnetic at low $T$ for $x \leq 0.045$. The $x =
0.065$ compound, on the other hand, shows a magnetic volume fraction
$V_{\textrm{m}} \simeq 0.7$ at $1.5$ K, namely the sample does not
fully turn magnetic down to the lowest investigated temperature. Any
further increase in $x$ is then expected to completely suppress
magnetism in the material.

\section{Superconducting phase}

\begin{figure}[b!]
\vspace{5cm} \includegraphics{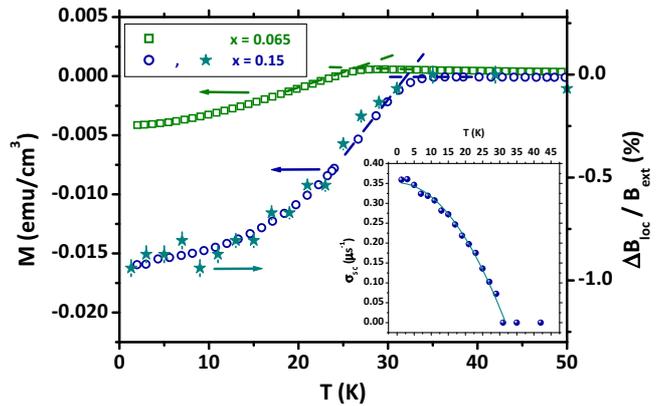} \caption{\label{SQUIDandMUSR} Main panel: FC
magnetization of the two samples $x = 0.065$ and $x = 0.15$ at $H =
1$ Oe (open symbols). Intercepts of the dashed lines allow the
estimate of $T_{\textrm{c}}$. Blue stars quantify the diamagnetic
screening probed by muons in $x = 0.15$ (see text). Inset:
$T$-dependence of the damping $\sigma_{\textrm{SC}}$ of the
TF-$\mu^{+}$SR signal introduced by the penetration of vortex lines
into the $x = 0.15$ sample (see text). The continuous line is a best
fit to data according to Eq. \ref{EqPowLawDensSup}.}
\end{figure}

Measurements of magnetic susceptibility were used to establish the
superconducting critical temperatures  $T_{\textrm{c}}$ for $x \geq
0.065$ samples. The results of field-cooled (FC) curves at $H = 1$
Oe are presented in Fig. \ref{SQUIDandMUSR}. For the two
investigated samples belonging to this doping region one finds
$T_{\textrm{c}}\left(x = 0.065\right) = 26.5 \pm 0.5$ K and
$T_{\textrm{c}}\left(x = 0.15\right) = 32.0 \pm 0.5$ K. By
extrapolating the magnetization values at zero $T$, it is possible
to compare the relative shielding fractions for the samples from the
ratio $M(0)_{x = 0.065}/M(0)_{x = 0.15} \sim 0.35$.

A more precise estimate of the absolute values of the shielding
fractions for the two samples is possible by considering the results
of the transverse-field (TF) $\mu^{+}$SR experiments performed on
the $x = 0.15$ sample at ISIS (Rutherford-Appleton Laboratories, UK)
in the MuSR spectrometer (raw data not shown). In this experiment a
magnetic field $|\textbf{H}_{\textrm{ext}}| = 150$ Oe is applied to
the sample in a direction perpendicular to the initial spin
polarization of the implanted muons. For $T \gg T_{\textrm{c}}$ the
sample is in a fully-paramagnetic regime and all the muons precess
coherently around $\textbf{H}_{\textrm{ext}}$. No damping of the
coherent precession is active, with the only exceptions of the weak
one associated to the nuclear magnetism, quantified by $\sigma$ as
in Eq. \ref{EqGeneralFittingZF}, and of the possible contribution
from diluted magnetic impurities, causing an exponential damping
quantified by $\lambda_{\textrm{imp}}$. On the other hand, a lattice
of vortex lines is expected to be present inside the sample for $T <
T_{\textrm{c}}$. This, in turn, gives rise to a modulation of the
spatial profile of $H_{\textrm{ext}}$ resulting in a Gaussian
extra-damping of the muon precession frequency \cite{Lue08}
quantified by $\sigma_{\textrm{SC}}$ in the fitting function
\begin{equation}\label{EqGeneralFittingTFNoMagnetism}
    A_{T}(t) = A_{0} \cos\left(\gamma B_{\mu} t\right)
    e^{-\frac{\left(\sigma_{\textrm{SC}}^{2} + \sigma^{2}\right)
    t^{2}}{2}} e^{-\lambda_{\textrm{imp}} t}.
\end{equation}
At the same time, the shielding of $H_{\textrm{ext}}$
by the superconducting phase leads to a sizeable lowering of the
internal field $B_{\mu}$ felt by the muons. This diamagnetic
shielding is typically quantified by the quantity $\Delta
B_{\textrm{loc}} \equiv B_{\mu} - H_{\textrm{ext}}$.

By fitting the experimental data above $T_{\textrm{c}}$ according to
Eq. \ref{EqGeneralFittingTFNoMagnetism}, one can find that the value
of $\lambda_{\textrm{imp}}$ is almost constant at $\sim 0.18 \;
\mu$s$^{-1}$ for $T_{\textrm{c}} \leq T \leq 100$ K. This allows us
to keep $\lambda_{\textrm{imp}}$ as a fixed parameter also for $T
\leq T_{\textrm{c}}$ incorporating all the extra-broadening of the
line in $\sigma_{\textrm{SC}}$. Fitting results are presented in
Fig. \ref{SQUIDandMUSR}. The $\sigma_{\textrm{SC}}$ vs $T$ trend is
well described by the function
\begin{equation}\label{EqPowLawDensSup}
    \sigma_{\textrm{SC}} = \sigma_{\textrm{SC}}(0) \left[1 -
    \left(\frac{T}{T_{\textrm{c}}}\right)^{2}\right]
\end{equation}
as previously reported for LaFeAsO$_{1-x}$F$_{x}$ \cite{Lue08}. The
saturation value $\sigma_{\textrm{SC}}(0) = 0.350 \pm 0.005 \;
\mu$s$^{-1}$ allows us to deduce an in-plane penetration depth
$\lambda_{\textrm{ab}}(0) = 420 \pm 10$ nm at zero $T$ (see Ref.
\cite{She12} for details).


From the observation that all the muons feel a static field lower
than $H_{\textrm{ext}}$ (Fig. \ref{SQUIDandMUSR}) we conclude that
superconductivity in the $x = 0.15$ compound is a bulk phenomenon
extended over the whole sample volume. Similarly, magnetization data
and TF-$\mu^{+}$SR measurements imply a superconducting volume
fraction of only $\sim 35 \%$ for the $x = 0.065$ compound.
Considering that the magnetic volume fraction at $T = 0$ K is $\sim
70 \%$ we argue that in this sample magnetism and superconductivity
are macroscopically separated. A similar scenario has been derived
in LaFeAsO$_{1-x}$F$_{x}$, where the ratio of the superconducting
and the magnetic volume fractions was tuned by the application of
external hydrostatic pressure \cite{Kha11}.

\section{Low-energy spin dynamics}

\begin{figure}[b!]
\vspace{6.2cm} \includegraphics{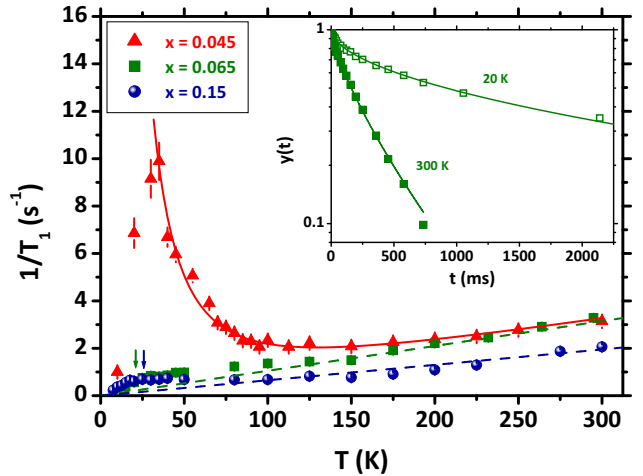} \caption{\label{GraRelRate} Main panel:
$1/T_{1}$ vs $T$ of {}$^{19}$F nuclei for the investigated samples.
The continuous line is a best fit to data according to Eq.
\ref{EqRelRate}. Dashed lines are linear best fit to data accounting
for Korringa-like relaxation. Arrows indicate the onset of
superconductivity for $x = 0.065$ and $x = 0.15$. Inset: $y(t)$ at
different $T$ for $x = 0.065$. Continuous lines are best fits to
data according to Eq. \ref{EqRelRateRecovery}. Measurements have
been performed at $H_{0} = 39$ KOe.}
\end{figure}

Further insights were obtained by means of {}$^{19}$F-NMR
measurements on the samples $0.045 \leq x \leq 0.15$. Here, the main
quantity of interest is the so-called spin-lattice relaxation time
$T_{1}$ for the {}$^{19}$F ($I = 1/2$) nuclear spins, quantifying
the time required by the nuclear magnetization $M$ to relax back to
the thermodynamical equilibrium along the quantization axis, once a
proper radio-frequency pulse sequence has brought the system into a
saturation condition \cite{Sli90}. A conventional sequence
$\left(\pi/2\right)_{\textrm{sat}} - t - \left(\pi/2\right) -
\tau_{\textrm{echo}} - \left(\pi/2\right)$ was employed to this aim.
The quantity
\begin{equation}\label{EqRelRateRecovery}
    y(t) = 1-\frac{M(t)}{M(\infty)} =
    \exp\left[-\left(\frac{t}{T_{1}}\right)^{\beta}\right]
\end{equation}
could then be measured, where the stretching parameter $\beta$
accounts for a distribution of relaxation times (raw data are
reported in the inset of Fig. \ref{GraRelRate}). When considering
relaxation processes involving nuclear spins with $I = 1/2$, values
$\beta < 1$ typically imply indeed an inhomogeneity of the local
environment of the nuclei. Typical measured values are $\beta \sim
0.6$ for $x = 0.045$ and $x = 0.065$ while $\beta \sim 0.85$ for $x
= 0.15$. The interplay between La$_{1-y}$Y$_{y}$ and
O$_{1-x}$F$_{x}$ dilutions is likely to be the main source of such
disorder in the examined compounds.

The relaxation rate $1/T_{1}$ as a function of $T$ is reported in
the main panel of Fig. \ref{GraRelRate} for the samples $x = 0.045$,
$x = 0.065$ and $x = 0.15$. By keeping into account the common
occurrence of power-law functional forms for $1/T_{1}$ in RE1111
materials \cite{Bru08,Pra10}, data relative to the fully magnetic
sample $x = 0.045$ can be properly described by the following
relation
\begin{equation}\label{EqRelRate}
    \frac{1}{T_{1}} =
    \left(\frac{1}{T_{1}}\right)_{\textrm{FL}} +
    \left(\frac{1}{T_{1}}\right)_{\textrm{SF}} =
    K_{\textrm{FL}} T + K_{\textrm{SF}} T^{-2}.
\end{equation}
where the exponent in the $\textrm{SF}$ contribution should be
considered as a phenomenological parameter. The term $\textrm{FL}$
accounts for relaxation processes driven by conduction electrons
belonging to a Fermi liquid and resulting in a linear $T$-dependence
(so-called Korringa-like relaxation) \cite{Sli90}. As a result of
the fitting procedure, $K_{\textrm{FL}} = 0.011 \pm 0.001$
s$^{-1}$K$^{-1}$. The term $\textrm{SF}$ in Eq. \ref{EqRelRate} can
be associated to spin fluctuations arising from the gradual magnetic
correlations eventually leading to the SDW phase at low $T$.

For the two superconducting samples the term $\textrm{FL}$ is the
leading relaxation channel besides small upturns with decreasing $T$
towards $T_{\textrm{c}}$. Fitting leads to $K_{\textrm{FL}} = 0.011
\pm 0.001$ s$^{-1}$K$^{-1}$ and $K_{\textrm{FL}} = 0.006 \pm 0.001$
s$^{-1}$K$^{-1}$ for $x = 0.065$ and $x = 0.15$, respectively,
suggesting a gradual suppression of the density of states at the
Fermi energy. With further decreasing of $T$ below $T_{\textrm{c}}$,
the relaxation rate in both samples decreases much more steeply.
This result again confirms that {}$^{19}$F nuclei are sensitive to
processes involving bands from the FeAs layers.

\section{Phase diagram and conclusions}

\begin{figure}[t!]
\vspace{6.2cm} \includegraphics{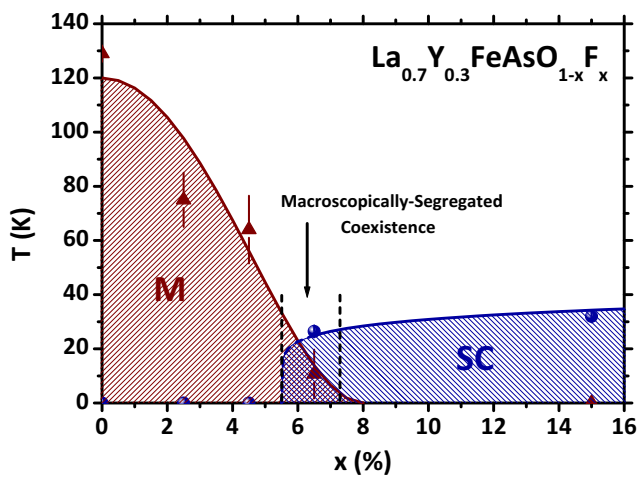} \caption{\label{PhDiag} Phase diagram for
the electronic ground states of
La$_{0.7}$Y$_{0.3}$FeAsO$_{1-x}$F$_{x}$ materials as a function of
$x$ as obtained by combining $\mu^{+}$SR and SQUID data. The
$T_{\textrm{N}}$ value for $x = 0$ is taken from resistivity
measurements reported in Ref. \cite{Mar09}.}
\end{figure}
The phase diagram in Fig. \ref{PhDiag} displays the magnetic and
superconducting transition temperatures as a function of $x$ for the
La$_{0.7}$Y$_{0.3}$FeAsO$_{1-x}$F$_{x}$ samples under investigation.

The main effect of F doping is the progressive suppression of
magnetism and the appearance of superconductivity, resembling the
general behavior of the RE1111 family. It is important to stress
that here the samples display macroscopic phase separation at the
magnetic-superconducting boundary. To the best of our knowledge,
this behavior has never been observed before for the RE1111
compounds at ambient pressure. Instead, materials with RE = La
display a sharp first-order-like magnetic-superconducting transition
and no coexistence has been reported up to now \cite{Lue09}.
However, macroscopic phase separation of magnetic and
superconducting volumes can be induced by applying hydrostatic
pressure to non-superconducting La1111 samples with F content close
to the disruption of magnetism \cite{Kha11}. By considering that a
similar macroscopic phase separation is here reported for
La$_{0.7}$Y$_{0.3}$FeAsO$_{1-x}$F$_x$ samples, we can conclude that
the chemical pressure produced by the partial Y substitution of La
mimics closely the effect of the hydrostatic pressure.


\begin{thebibliography}{[1]}
\bibitem{San10}%
 S. Sanna, R. De Renzi, T. Shiroka, G. Lamura, G. Prando, P. Carretta,
 M. Putti, A. Martinelli, M. R. Cimberle, M. Tropeano, and A.
 Palenzona, Phys. Rev. \textbf{B 82}, 060508(R) (2010).
\bibitem{San11}%
 S. Sanna, P. Carretta, P. Bonf\'a, G. Prando, G. Allodi,
 R. De Renzi, T. Shiroka, G. Lamura, A. Martinelli, and M. Putti,
 Phys. Rev. Lett. \textbf{107}, 227003 (2011).
\bibitem{Wie11}%
 E. Wiesenmayer, H. Luetkens, G. Pascua, R. Khasanov, A. Amato,
 H. Potts, B. Banusch, H.-H. Klauss, and D. Johrendt, Phys. Rev.
 Lett. \textbf{107}, 237001 (2011).
\bibitem{Shi11}%
 T. Shiroka, G. Lamura, S. Sanna, G. Prando, R. De
 Renzi, M. Tropeano, M. R. Cimberle, A. Martinelli,
 C. Bernini, A. Palenzona, R. Fittipaldi, A. Vecchione, P.
 Carretta, A. S. Siri, C. Ferdeghini, and M. Putti, Phys.
 Rev. B \textbf{84}, 195123 (2011).
\bibitem{She12}%
 Z. Shermadini, H. Luetkens, R. Khasanov, A. Krzton-Maziopa, K.
 Conder, E. Pomjakushina, H.-H. Klauss, and A. Amato,
 Phys. Rev. B \textbf{85}, 100501(R) (2012).
\bibitem{Jul09}%
 M.-H. Julien, H. Mayaffre, M. Horvati\'c, C. Berthier, X. D. Zhang,
 W. Wu, G. F. Chen, N. L. Wang and J. L. Luo,
 Europhys. Lett. \textbf{87}, 37001 (2009).
\bibitem{Tex12}%
 Y. Texier, J. Deisenhofer, V. Tsurkan, A. Loidl, D. S. Inosov, G.
 Friemel, and J. Bobroff, Phys. Rev. Lett. \textbf{108}, 237002
 (2012).
\bibitem{Lue09}%
 H. Luetkens, H.-H. Klauss, M. Kraken, F. J. Litterst, T. Dellmann,
 R. Klingeler, C. Hess, R. Khasanov, A. Amato, C. Baines, M. Kosmala,
 O. J. Schumann, M. Braden, J. Hamann-Borrero, N. Leps, A. Kondrat,
 G. Behr, J. Werner, and B. B\"uchner, Nature Mater.
 \textbf{8}, 305 (2009).
\bibitem{Kha11}%
 R. Khasanov, S. Sanna, G. Prando, Z. Shermadini, M. Bendele, A. Amato,
 P. Carretta, R. De Renzi, J. Karpinski, S. Katrych, H. Luetkens, and
 N. D. Zhigadlo, Phys. Rev. B \textbf{84}, 100501(R) (2011).
\bibitem{Tro09}%
 M. Tropeano, C. Fanciulli, F. Canepa, M. R. Cimberle, C. Ferdeghini,
 G. Lamura, A. Martinelli, M. Putti, M. Vignolo, and A. Palenzona,
 Phys. Rev. B \textbf{79}, 174523 (2009).
\bibitem{Mar09}%
 A. Martinelli, A. Palenzona, M. Tropeano, C. Ferdeghini, M. R. Cimberle,
 and C. Ritter, Phys. Rev. B \textbf{80}, 214106 (2009).
\bibitem{DeR12}%
 R. De Renzi, P. Bonf\'a, M. Mazzani, S. Sanna, G. Prando, P.
 Carretta, R. Khasanov, A. Amato, H. Luetkens, M. Bendele, F.
 Bernardini, S. Massidda, A. Palenzona, M. Tropeano, and M. Vignolo,
 Supercond. Sci. Technol. \textbf{25}, 084009 (2012).
\bibitem{Lue08}%
 H. Luetkens, H.-H. Klauss, R. Khasanov, A. Amato, R. Klingeler, I.
 Hellmann, N. Leps, A. Kondrat, C. Hess, A. K\"ohler, G. Behr, J. Werner,
 and B. B\"uchner, Phys. Rev. Lett. \textbf{101}, 097009 (2008).
\bibitem{Sli90}%
 C. P. Slichter, {\it Principles of Magnetic Resonance},
 Springer-Verlag Berlin (1990)
\bibitem{Bru08}%
 E. M. Br\"uning, C. Krellner, M. Baenitz A. Jesche, F. Steglich,
 and C. Geibel, Phys. Rev. Lett. \textbf{101}, 117206 (2008).
\bibitem{Pra10}%
 G. Prando, P. Carretta, A. Rigamonti, S. Sanna, A. Palenzona, M. Putti,
 and M. Tropeano, Phys. Rev. B \textbf{81}, 100508(R) (2010).
\end{thebibliography}
\end{document}